\documentclass[12pt,a4paper,final,nofootinbib]{revtex4}

\usepackage{sidecap}

\usepackage{ulem}
\usepackage{epsfig}
\usepackage{amsmath,amssymb,amsthm,mathtools}
\usepackage{graphicx}
\usepackage{bm,url,enumitem}
\usepackage{color,soul}
\setstcolor{blue}
\sethlcolor{blue}
\DeclareMathAlphabet{\mathpzc}{OT1}{pzc}{m}{it}

\begin{document}

\renewcommand{\textfraction}{0.00}

% Useful macros:

\newcommand{\pT}{p_\perp}
\newcommand{\vAi}{{\cal A}_{i_1\cdots i_n}}
\newcommand{\vAim}{{\cal A}_{i_1\cdots i_{n-1}}}
\newcommand{\vAbi}{\bar{\cal A}^{i_1\cdots i_n}}
\newcommand{\vAbim}{\bar{\cal A}^{i_1\cdots i_{n-1}}}
\newcommand{\htS}{\hat{S}}
\newcommand{\htR}{\hat{R}}
\newcommand{\htB}{\hat{B}}
\newcommand{\htD}{\hat{D}}
\newcommand{\htV}{\hat{V}}
\newcommand{\cT}{{\cal T}}
\newcommand{\cM}{{\cal M}}
\newcommand{\cMs}{{\cal M}^*}
\newcommand{\vk}{\vec{\mathbf{k}}}
\newcommand{\bk}{\bm{k}}
\newcommand{\kt}{\bm{k}_\perp}
\newcommand{\kp}{k_\perp}
\newcommand{\km}{k_\mathrm{max}}
\newcommand{\vl}{\vec{\mathbf{l}}}
\newcommand{\bl}{\bm{l}}
\newcommand{\bK}{\bm{K}}
\newcommand{\bb}{\bm{b}}
\newcommand{\qm}{q_\mathrm{max}}
\newcommand{\vp}{\vec{\mathbf{p}}}
\newcommand{\bp}{\bm{p}}
\newcommand{\vq}{\vec{\mathbf{q}}}
\newcommand{\bq}{\bm{q}}
\newcommand{\qt}{\bm{q}_\perp}
\newcommand{\qp}{q_\perp}
\newcommand{\bQ}{\bm{Q}}
\newcommand{\vx}{\vec{\mathbf{x}}}
\newcommand{\bx}{\bm{x}}
\newcommand{\tr}{{{\rm Tr\,}}}
\newcommand{\sNN}{s_{\mathrm{NN}}}
\newcommand{\bc}{\textcolor{blue}}
\newcommand{\lton}{\mathrel{\lower.9ex \hbox{$\stackrel{\displaystyle <}{\sim}$}}}
\newcommand{\tcr}{\textcolor{red}}
\newcommand{\beq}{\begin{equation}}
\newcommand{\eeq}[1]{\label{#1} \end{equation}}
\newcommand{\ee}{\end{equation}}
\newcommand{\bea}{\begin{eqnarray}}
\newcommand{\eea}{\end{eqnarray}}
\newcommand{\beqar}{\begin{eqnarray}}
\newcommand{\eeqar}[1]{\label{#1}\end{eqnarray}}

\newcommand{\dif}{\mathrm{d}}
\newcommand{\RAA}{$R_{AA}$}

\newcommand{\half}{{\textstyle\frac{1}{2}}}
\newcommand{\ben}{\begin{enumerate}}
\newcommand{\een}{\end{enumerate}}
\newcommand{\bit}{\begin{itemize}}
\newcommand{\eit}{\end{itemize}}
\newcommand{\ec}{\end{center}}
\newcommand{\bra}[1]{\langle {#1}|}
\newcommand{\ket}[1]{|{#1}\rangle}
\newcommand{\norm}[2]{\langle{#1}|{#2}\rangle}
\newcommand{\brac}[3]{\langle{#1}|{#2}|{#3}\rangle}
\newcommand{\hilb}{{\cal H}}
\newcommand{\pleft}{\stackrel{\leftarrow}{\partial}}
\newcommand{\pright}{\stackrel{\rightarrow}{\partial}}

\newcommand{\meqbox}[2]{\eqmakebox[#1]{$\displaystyle#2$}}
%%%%%%%%%%%%%%%%%%%%%%%%%%%%%%%%%%%%%%%%%%%%%%%%%%%%%%%%%%%%%%%%%%%%%%%%%%%%%%

\title{Constraining $\eta /s$ through high-p$_\perp$ theory and data}

\author{Bithika Karmakar}
\affiliation{Institute of Physics Belgrade, University of Belgrade, Belgrade 11080, Serbia}

\author{Dusan Zigic}
\affiliation{Institute of Physics Belgrade, University of Belgrade, Belgrade 11080, Serbia}

\author{Igor Salom}
\affiliation{Institute of Physics Belgrade, University of Belgrade, Belgrade 11080, Serbia}

\author{Jussi Auvinen}
\affiliation{Institute of Physics Belgrade, University of Belgrade, Belgrade 11080, Serbia}
\affiliation{University of Jyv\"{a}skyl\"{a}, Jyv\"{a}skyl\"{a} P.O. Box 35, FI-40014, Finland}

\author{Pasi Huovinen}
\affiliation{Incubator of Scientific Excellence---Centre for Simulations of Superdense Fluids, University of Wroc\l{}aw, Wroc\l{}aw 50-204, Poland  }

\author{Marko Djordjevic}
\affiliation{Faculty of Biology, University of Belgrade, Belgrade 11000, Serbia}

\author{Magdalena Djordjevic\footnote{E-mail: magda@ipb.ac.rs}}
\affiliation{Institute of Physics Belgrade, University of Belgrade, Belgrade 11080, Serbia}

\begin{abstract}
We study whether it is possible to use high-$\pT$ data/theory to
constrain the temperature dependence of the shear viscosity over
entropy density ratio $\eta/s$ of the matter formed in
ultrarelativistic heavy-ion collisions at the BNL Relativistic Heavy Ion Collider (RHIC) and the CERN Large Hadron Collider (LHC). We use two
approaches: {\it i}) We calculate high-$p_\perp$ $R_{AA}$ and flow
coefficients $v_2$, $v_3$ and $v_4$ assuming different $(\eta/s)(T)$
of the fluid-dynamically evolving medium. {\it ii}) We calculate the
quenching strength ($\hat{q}/T^3$) from our dynamical energy loss
model and convert it to $\eta/s$ as a function of temperature. It
turned out that the first approach can not distinguish between
different $(\eta/s)(T)$ assumptions when the evolution is constrained
to reproduce the low-$\pT$ data. In distinction, $(\eta/s)(T)$ calculated using the second approach
agrees surprisingly well with the $(\eta/s)(T)$ inferred through
state-of-the-art Bayesian analyses of the low-$\pT$ data even in the
vicinity of $T_c$, while providing much smaller uncertainties at high
temperatures.
\end{abstract}

\pacs{12.38.Mh; 24.85.+p; 25.75.-q}
\maketitle

\section{Introduction}

Quantum chromodynamics (QCD) predicts that at extremely high densities
matter undergoes a transition to a state consisting of deconfined and
interacting quarks, antiquarks, and gluons~\cite{Collins,Baym}.
According to the current cosmology, this new state of matter, called
Quark-Gluon plasma (QGP)~\cite{Shuryak}, existed immediately after the
Big Bang~\cite{Stock}. Today, QGP is created in $``$Little Bangs",
when heavy ions collide at ultra-relativistic energies~\cite{GML} in experiments at the BNL Relativistic	
Heavy Ion Collider (RHIC) and the CERN Large Hadron	
Collider (LHC). Such collisions lead to an expanding
fireball of quarks and gluons, which thermalizes to form QGP. The QGP
cools down, and quarks and gluons hadronize when the temperature $T$
drops to the critical temperature $T_c$.

Extracting useful information from $``$Little Bangs" requires
comparing theoretical predictions with experimental data. By such
comparisons, it is established that QGP is formed at the RHIC and LHC
experiments~\cite{JS} through two main lines of
evidence~\cite{GML,JS,Stachel}: i) by comparison of low transverse
momentum ($p_\perp$) measurements with relativistic hydrodynamical
predictions, which imply that created QGP is consistent with the
description of a nearly perfect
fluid~\cite{KolbHeinz,Romatschke,HeinzSnellings}, ii) by comparison of perturbative QCD (pQCD) predictions with high-$p_\perp$ data~\cite{Adams,Adcox,Aad,Aamodt},
which showed that high-$p_\perp$ partons (jets) significantly interact
with the opaque medium~\cite{GML}. Beyond the discovery phase, the
current challenge is to investigate the properties of this extreme
form of matter~\cite{Novak:2013bqa,Pratt:2015zsa,Bernhard:2018hnz,Bernhard:2019bmu,Auvinen:2020mpc,JETSCAPE:2020mzn,JETSCAPE:2020shq,Stojku:2020wkh,Stojku:2021yow,Soloveva:2021quj,Xing:2023ciw}.

The QGP was expected to behave as a weakly interacting gas based on
ideas of asymptotic freedom and color screening~\cite{JohSt}. Thus,
the agreement of the fluid-dynamical predictions, which assumed the
QGP to behave as a nearly inviscid fluid, with the data came as a
surprise~\cite{HeinzSnellings}. Furthermore, subsequent calculations
revealed~\cite{HeinzSnellings} that reproduction of the data required
the shear viscosity to entropy density ratio ($\eta/s$) of QGP to be
near the lower bound predicted by anti–de Sitter	
and conformal field theory (AdS/CFT) correspondence~\cite{Kovtun}.

However, the temperature of the QGP changes
significantly~\cite{Heinz:2015tua} during the evolution of the
collision system. E.g., in the LHC experiments, the temperature is
estimated to span the range from $4T_c$ to $T_c$. Even if the QGP
behaves as a perfect fluid close to $T_c$ (the $``$soft", strongly
coupled, regime), its $\eta/s$ may significantly increase with
increasing $T$ if the QGP becomes weakly coupled at higher
temperatures (the $``$hard", weakly coupled, regime). We call this
possibility the $``$soft-to-hard" medium hypothesis.

Testing this hypothesis has turned out to be surprisingly
difficult. Reproduction of the observed anisotropies of low-$\pT$
particles necessitates low $\eta/s$ in the vicinity of $T_c$, but the
value of shear viscosity in higher temperatures has only a weak effect
on anisotropies in collisions at LHC energies, and basically no effect
at all at RHIC~\cite{Niemi:2011ix,Nagle,Niemi:2012ry}. In the recent
Bayesian analyses of the data, this is manifested in a well
constrained $\eta/s$ in the $T_c \lton T \lton 1.5T_c$ temperature
range and weak constraints at larger temperatures~\cite{Bernhard:2018hnz,
  Bernhard:2019bmu,Auvinen:2020mpc,JETSCAPE:2020mzn}. Some of the most recent Bayesian analyses~\cite{Nijs:2022rme,Heffernan:2023utr} even suggest that $\eta/s$ may decrease in the region $T_c - 2T_c$, where the reason for such a decrease still remain to be understood.

Thus, it is evident that a complementary theory and observables are
needed to investigate the $``$soft-to-hard" medium hypothesis. Since
most of the jet energy loss takes place when the system is hottest, it
is reasonable to expect the high-$\pT$ observables to be sensitive to
the properties of the system at that stage. To use jet energy loss and
high-$\pT$ data to provide constraints to the bulk properties of the
collision system, we developed the state-of-the-art DREENA tomography
tool~\cite{Zigic:2021rku,Zigic:2022xks} based on the dynamical energy
loss formalism~\cite{Djordjevic:2006tw,Djordjevic:2009cr,Djordjevic:2008iz}.
So far, we have used this tool to, e.g., provide constraints to the
early evolution of the collision system~\cite{Stojku:2020wkh} and map
how the shape of the collision system is manifested in the high-$\pT$
data~\cite{Stojku:2021yow}.

In this study, we explore whether high-$\pT$ data can provide
constraints on the $\eta/s$ ratio of QGP at high temperatures. As
known, shear viscosity generates entropy, which means that the system
with larger viscosity cools slower or, alternatively, to reach the
same final entropy, the system with larger viscosity must have a lower
initial temperature. Thus, different assumed $\eta/s$ during the early
evolution of the system may lead to different jet energy loss and
therefore different nuclear suppression factor $R_{AA}$. As well,
azimuthal anisotropy in path lengths and temperature along the paths
leads to azimuthal dependence of jet suppression~\cite{GML}, which is
measured as $v_n$ of high-$\pT$ particles. High-$\pT$ $v_n$ are known
to be sensitive to the details of the medium
evolution~\cite{Renk,Zigic:2021rku, Zigic:2022xks}, and since
viscosity changes the evolution of the anisotropy of the system, the
changes in $\eta/s$ can lead to changes in high-$\pT$ $v_n$. We choose
three different parametrizations of $(\eta/s)(T)$, adjust the
parameters to reproduce the low-$\pT$ data measured in $\sqrt{\sNN} =
200$ GeV Au+Au (RHIC) and $\sqrt{\sNN} = 5.02$ TeV Pb+Pb collisions
(LHC), calculate the temperature evolution of the system, and energy
loss of jets traversing this system in each case, and evaluate the
$R_{AA}$ and high-$\pT$ $v_2$, $v_3$ and $v_4$ to see if different
assumptions of $\eta/s$ lead to differences in these observables.

Complementary to this phenomenological approach to infer the $\eta/s$
ratio from the experimental data, we also provide a fully theoretical
estimate of $\eta/s$ based on jet energy loss: The jet quenching
strength is quantified through the jet quenching parameter
$\hat{q}$. It has been argued that in a weakly coupled regime
$T^3/\hat{q}$ is directly proportional to $\eta/s$~\cite{MBW}, and
thus evaluating one allows one to know the other. We estimate the
quenching parameter $\hat{q}$ as function of temperature using our
dynamical energy loss formalism, convert it to $\eta/s$ and compare
the resulting $(\eta/s)(T)$ to constraints obtained from
state-of-the-art Bayesian
analyses~\cite{Bernhard:2019bmu,Auvinen:2020mpc}.

\section{Methods}

\subsection{Modeling the bulk evolution}
   \label{hydro}

To calculate the temperature evolution and the low-$\pT$ observables
we use the version of VISHNew~\cite{Song:2007ux,Song:2008si} used in
Refs.~\cite{Bernhard:2016tnd,Bernhard:2018hnz,Bernhard:2019bmu}\footnote{Code
  available at~\cite{Jonahs_github}}. It is a code to solve the
dissipative fluid-dynamical equations in 2+1-dimensions, i.e.,
assuming boost invariance. Shear stress and bulk pressure are taken as
dynamical variables and evolved according to the Israel-Stewart type
equations~\cite{Israel:1979wp}. We use an Equation of State
(EoS)~\cite{Bernhard:2016tnd} that combines the lattice QCD-based EoS
of the HotQCD collaboration~\cite{HotQCD:2014kol} at large
temperatures to a hadron resonance gas EoS at low temperatures. At
constant temperature, $T_{\mathrm{sw}} = 151$ MeV, hypersurface we
convert the fluid to particle ensembles according to the Cooper-Frye
prescription~\cite{Cooper:1974mv}. These ensembles are fed to the
UrQMD hadron cascade~\cite{Bass:1998ca,Bleicher:1999xi}, which
describes the evolution of the hadronic stage of the system until
freeze-out.

We generate the event-by-event fluctuating initial states using the
T$_{\text{R}}$ENTo model~\cite{Moreland:2014oya}. In this model
nucleus-nucleus collisions are considered as a superposition of
nucleon-nucleon collisions. The nucleons are represented by Gaussian
distributions, which in this study have the width $w=0.5$ fm while the
minimum nucleon-nucleon distance within the nucleus is also set to
$d=0.5$ fm. The inelastic nucleon-nucleon cross section is $70$ mb at
$\sqrt{s}=5.02$ TeV (energy of Pb+Pb collisions) and $42$ mb at
$\sqrt{s}=200$ GeV (energy of Au+Au collisions). For the other parameters
we use the maximum a posteriori (MAP) values found in
Ref.~\cite{Bernhard:2019bmu}. We do not allow any pre-equilibrium
evolution (free-streaming or otherwise), and use $\tau_0 = 1$ fm/$c$
as the initial time for fluid-dynamical evolution, since the
reproduction of the high-$\pT$ observables does not allow strong
transverse expansion earlier~\cite{Stojku:2020wkh}.

We include both bulk and shear viscosity in our fluid-dynamical
calculation. The temperature dependence of the bulk viscosity
coefficient $\zeta$ is parameterized as a Cauchy
distribution~\cite{Bernhard:2019bmu}:
\bea
  (\zeta/s)(T) = \frac{(\zeta/s)_{\text{max}}}
                      {1+\Big( \frac{T-T_0}{(\zeta/s)_{\text{width}}} \Big)^2}.
\eea
We consider a small bulk viscosity with a maximum value
$(\zeta/s)_{\text{max}}$ = 0.03, the width parameter
$(\zeta/s)_{\text{width}}$ = 0.022 and $T_0$ = 0.183 GeV. As in the
case of the T$_{\text{R}}$ENTo parameters described above, the width
and $T_0$ correspond to MAP parameter values from
Ref.~\cite{Bernhard:2019bmu}. However, the maximum of the bulk viscosity ($(\zeta/s)_{\text{max}}$) is decreased compared to the MAP value in~\cite{Bernhard:2019bmu} to compensate the lack of pre-equilibrium free streaming and still reach agreement with the $p_\perp$ spectra.

As mentioned, our main objective is to find out whether high-$\pT$
data can provide constraints to $\eta/s$ at high temperatures. Naively
one can expect the jet energy loss to be proportional to the third
power of temperature ($T^3$), but a detailed calculation has shown it
to be proportional to only $T^{1.2}$~\cite{Djordjevic:2015hra,
  Djordjevic:2019tdu}. Since the sensitivity to temperature is weaker
than expected, we want to maximize the difference in temperature due
to differences in $\eta/s$. Therefore we do not take as our
$(\eta/s)(T)$ the upper and lower limits suggested by the Bayesian
analyses~\cite{Bernhard:2019bmu,Auvinen:2020mpc} but something more
extreme. We parameterize the temperature dependence of $\eta/s$
as~\cite{Bernhard:2018hnz,Bernhard:2019bmu}
\bea
 (\eta/s)(T) = \left\{
   \begin{aligned}
      &(\eta/s)_{\text{min}}, &&  T<T_c, \\
      &(\eta/s)_{\text{min}} + (\eta/s)_{\text{slope}} (T-T_c)
                        \Big( \frac{T}{T_c} \Big)^{(\eta/s)_{\text{crv}}}, &&  T>T_c,
   \end{aligned}\right.
\eea
where $(\eta/s)_{\text{min}}$ is the minimum value of the specific
shear viscosity, $(\eta/s)_{\text{slope}}$ is the slope above $T_c$
and $(\eta/s)_{\text{crv}}$ controls the curvature above $T_c$. $T_c$
is fixed to the pseudocritical temperature $T_c = 154$ MeV evaluated
by the HotQCD collaboration~\cite{HotQCD:2014kol}.

We study three different scenarios, each capable of describing a
subset of low-$\pT$ data at RHIC and LHC with reasonable accuracy:
\begin{enumerate}[label=(\roman*),noitemsep,topsep=0pt]
  \item constant $\eta/s$ (0.15 for Pb+Pb collision at LHC and 0.12
    for Au+Au collision at RHIC)
  \item $(\eta/s)_{\text{min}}$ = 0.1, $(\eta/s)_{\text{slope}}$= 1.11,
    $(\eta/s)_{\text{crv}}$ = -0.48,
  \item $(\eta/s)_{\text{min}}$ = 0.04, $(\eta/s)_{\text{slope}}$ = 3.30,
    $(\eta/s)_{\text{crv}}$ = 0.
\end{enumerate}
The parameters in our second scenario are within the 90\% credible
intervals of the analysis of Ref.~\cite{Bernhard:2019bmu}.
Therefore, we label it as `Nature'. Nevertheless, our
$(\eta/s)_{\text{min}}$ is larger than in Ref.~\cite{Bernhard:2019bmu}
since we require the reproduction of the RHIC data, not only the
LHC data. As known, including the RHIC data tends to
increase the favored minimum value of $\eta/s$~\cite{JETSCAPE:2020mzn}.
Our third scenario with its very rapidly rising $\eta/s$ (see
Fig.~\ref{Panel}) is inspired by the `LHHQ' parametrization in
Ref.~\cite{Niemi:2011ix}. Consequently, we label it as such.

To calculate the low-$\pT$ and high-$\pT$ predictions, we generated $10^4$ minimum-bias events and sorted the events in centrality classes according to the number of participants. While using the final particle multiplicity would be closer to the centrality selection done in experiments, participant number sorting allows us to reduce the number of hydrodynamic simulations by focusing on the narrower (10-50)\% centrality range, thus saving computational resources (we numerically tested that this approximation would have a negligible effect on theoretical predictions). Finally, we evaluated the event-averaged observables in each centrality bin.

\begin{center}
\begin{figure}[tbh!]
 \includegraphics[scale=0.5]{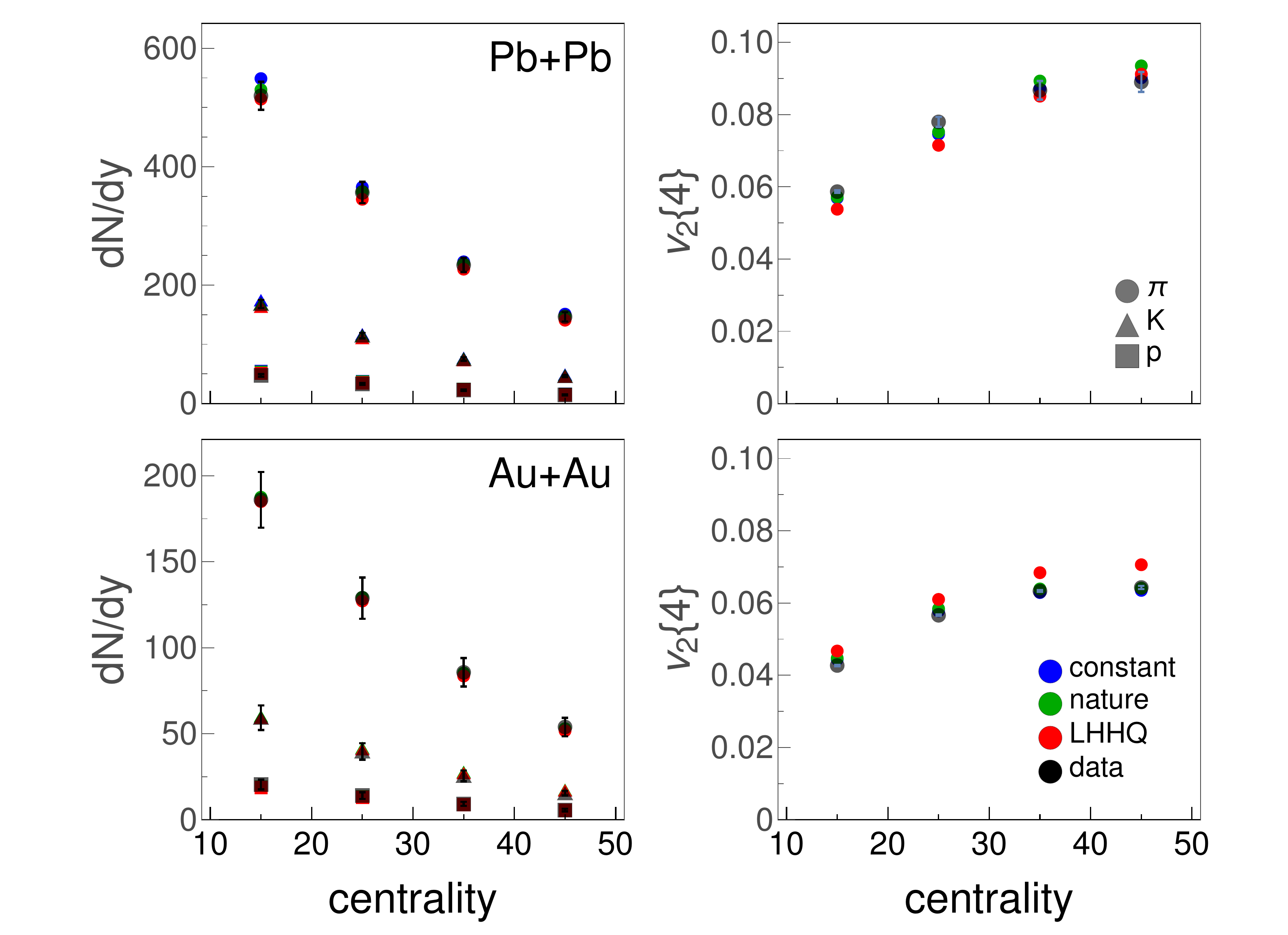}
 \caption{
 Left panels: Centrality dependence of the $p_\perp$-integrated yields of pions, kaons, and (anti-protons) are shown in different centrality classes. The pion multiplicity is scaled by 0.5. The upper panel corresponds to 5.02 TeV Pb+Pb collisions, where the ALICE experimental data is taken from Ref.~\cite{ALICE:2019hno}. The lower panel corresponds to 200 GeV Au+Au collisions, where the PHENIX experimental data is taken from Ref.~\cite{PHENIX:2003iij}. Right panels: $v_2\{4\}$ is shown at different centrality classes. The upper panel corresponds to 5.02 TeV Pb+Pb collisions, where the ALICE experimental data is taken from Ref.~\cite{ALICE:2016ccg}. The lower panel corresponds to 200 GeV Au+Au collisions, where the STAR experimental data is taken from  Ref.~\cite{STAR:2004jwm}.}
  \label{N_v24}
\end{figure}
\end{center}

We reproduced the pion, kaon, and proton multiplicities and charged
hadron 4-particle cumulant elliptic flow $v_2\{4\}$ in Au+Au
collisions at $\sqrt{\sNN} = 200$ GeV (RHIC) and Pb+Pb collisions at
$\sqrt{\sNN} = 5.02$ TeV (LHC) in 10--20\%, 20--30\%, 30--40\% and
40--50\% centrality classes by varying only the nucleon-nucleon cross
section according to the collision energy and the overall
normalization factor according to the collision energy and choice of
$(\eta/s)(T)$. All the other T$_{\text{R}}$ENTo parameters were kept
the same in all cases. For LHHQ parametrization, the minimum value of
$\eta/s$ is chosen to get an acceptable agreement of $v_2\{4\}$ with
both Pb+Pb and Au+Au collision data. The centrality dependence of
charged particle multiplicities ($p_\perp$-integrated yields) for
Pb+Pb and Au+Au collisions with three different $(\eta/s)(T)$
parametrizations found from the hydrodynamical simulation are shown in
the left panels of Fig.~\ref{N_v24}. The fourth-order cumulant of the
elliptic flow coefficient $v_2\{4\}$ at different centrality classes
for Pb+Pb and Au+Au collisions are shown in the right panels of
Fig.~\ref{N_v24}.

\subsection{Overview of DREENA framework}

After evaluating the temperature evolution, we use the $`$generalized
DREENA-A' framework to calculate the high-$\pT$ observables: nuclear
suppression factor $R_{AA}$ and high-$p_\perp$ flow harmonics $v_2$,
$v_3$ and $v_4$.
$`$DREENA' (Dynamical Radiative and Elastic ENergy loss Approach) is a
computationally efficient tool for QGP tomography~\cite{Zigic:2022xks,
  Zigic:2021rku}, based on generalized hard thermal loop (HTL)
perturbation theory~\cite{Kapusta:1989tk} with naturally regulated
infrared divergences~\cite{Djordjevic:2006tw,Blagojevic:2018nve}. In
this formalism both the radiative~\cite{Djordjevic:2009cr,Djordjevic:2008iz}
and collisional energy loss~\cite{Djordjevic:2006tw} of high energy
particles have been computed in an evolving QCD medium of finite size
at finite temperature. Furthermore, the framework is extended to
account for running coupling~\cite{Djordjevic:2013xoa}, finite
magnetic mass~\cite{Djordjevic:2011dd} and beyond soft-gluon
approximation~\cite{Blagojevic:2018nve}. Recently extended the formalism towards finite orders in opacity~\cite{Stojku:2023ell}, but showed that higher-order effects can be neglected for high-$p_\perp$ predictions. Thus, a computationally more efficient version with one scattering center is used in this study. Additionally, in this framework, all parameters are fixed to standard literature values stated below (i.e., no fitting parameters are used)~\cite{Stojku:2021yow,Stojku:2020wkh}. This allows systematic comparison of data and the predictions from the simulation obtained using the same formalism and parameter set.

We use the generic pQCD convolution formula~\cite{Djordjevic:2013xoa,
  Wicks:2005gt} to generate the final quenched ($q$) and unquenched ($u$)
spectra of hadrons as
\bea
\frac{E_f \dif^3\sigma_q(H_Q)}{\dif p^3_f}=\frac{E_i\dif^3\sigma(Q)}{\dif p^3_i}\otimes P(E_i \rightarrow E_f) \otimes D(Q\rightarrow H_Q),
\eea

\bea
\frac{E_f \dif^3\sigma_u(H_Q)}{\dif p^3_f}=\frac{E_i\dif^3\sigma(Q)}{\dif p^3_i}\otimes D(Q\rightarrow H_Q),
\eea
where $i$ and $f$ denote the initial parton $(Q)$ and the final hadron
$(H_Q)$ respectively. $\frac{E_i\dif^3\sigma(Q)}{\dif p^3_i}$ represents the
initial parton spectrum calculated at the next-to-leading order for
light and heavy partons~\cite{Kang:2012kc, Sharma:2009hn, Cacciari:2012ny}.
$P(E_i \rightarrow E_f)$ is the energy loss probability computed within
finite temperature field theory. $D(Q\rightarrow H_Q)$ represents the
fragmentation function. DSS~\cite{deFlorian:2007aj},
BCFY~\cite{Cacciari:2003zu, Braaten:1994bz} and
KLP~\cite{Kartvelishvili:1977pi} fragmentation functions have been
used for charged hadrons, D mesons, and B mesons, respectively. $`$DREENA-A'~\cite{Zigic:2021rku}, where $`$A' stands for
Adaptive ({\it{i.e.}}, Arbitrary) temperature profiles has been
optimized to incorporate any event-by-event fluctuating
temperature profile~\cite{Zigic:2022xks}. For parameters, we
use $\Lambda_{QCD} = 0.2$ GeV~\cite{Peshier:2006ah,Cao:2020wlm,Nakamura_LQCD} and the effective number of light
quark flavors $n_f=3$ and 2.5 for Pb+Pb and Au+Au collision systems,
respectively. We also consider the gluon mass
$m_g=\mu_E/\sqrt{2}$~\cite{Djordjevic:2003be} where $\mu_E$ is the
temperature-dependent Debye mass computed following the procedure in
Ref.~\cite{Peshier:2006ah} (outlined in the next subsection). We assume the mass of the light quark,
charm, and bottom quark to be $M=\mu_E/6$, 1.2 GeV, and 4.75 GeV,
respectively. The magnetic-to-electric mass ratio is
$\mu_M/\mu_E=0.6$~\cite{Borsanyi:2015yka}.

\subsection{Derivation of transport coefficient $\hat{q}$ from dynamical
  energy loss formalism}
    \label{derive-qhat}

To derive the transport coefficient $\hat{q}$,  which is the squared
 average transverse momentum exchange between the medium and the
 fast parton per unit path length~\cite{BDMPS}, we start from dynamical
{\it perturbative} QCD medium, where the interaction between
high-$p_\perp$ partons and QGP constituents can be characterized by
the HTL resummed elastic collision rate~\cite{JET:2013cls}:
\bea
\frac{\dif \Gamma_{el}}{\dif^2 q}=4 C_A\left(1+\frac{n_f}{6}\right) T^3 \frac{\alpha_s^2}{q^2\left(q^2+\mu_E^2\right)}. \label{col_rate}
\eea

While $\alpha_s$ in Eq.~\eqref{col_rate} is presumed to be constant,
for RHIC and LHC, it is necessary to include running coupling constant
in the kernel due to the wide kinematic range covered in these
experiments. To include running coupling in dynamical energy loss
formalism, we adopt the procedure from Ref.~\cite{Peigne-Peshier}
where
\bea
\alpha_s^2 \rightarrow \alpha_s(E T) \alpha_s\left(\mu_E^2\right) \label{alpha_change}
\eea
and $\mu_E$ is obtained~\cite{Peshier:2006ah} as a self-consistent solution to
%by self consistently solving the equation~\cite{Peshier:2006ah} as
\bea
\mu_E^2=\left(1+\frac{n_f}{6}\right) 4 \pi \alpha\left(\mu_E^2\right) T^2,
\eea
where
\bea
\alpha(t)=\frac{4\pi}{(11-\frac{2}{3}n_f  )}\frac{1}{\ln{(\frac{t}{\Lambda^2})}}, \label{running_coupling}
\eea
leading to
\bea
\mu_E=\sqrt{\Lambda^2 \frac{\xi(T)}{W(\xi(T))}}, \label{Debye_mass}
\eea
where
\bea
\xi(T)=\frac{1+\frac{n_f}{6}}{11-\frac{2}{3} n_f} \left(\frac{4 \pi T}{\Lambda}\right)^2 , \label{xi_T}
\eea
and $W$ is Lambert's $W$ function. Note that $\mu_E$ obtained through this procedure agrees with Lattice QCD results~\cite{Peshier:2006ah}.

By using Eq.~\eqref{Debye_mass} and \eqref{alpha_change}, Eq.~\eqref{col_rate} reduces to
\bea
\frac{\dif \Gamma_{el}}{\dif^2 q}=\frac{C_A}{\pi} T \frac{\alpha(E T)\mu_E^2}{q^2(q^2+\mu_E^2)}, \label{col_rate_2}
\eea
which reproduces Eq.~(16) from Ref.~\cite{JET:2013cls} in the case of
constant coupling $\alpha(ET)=g^2/4\pi$.  Following
Ref.~\cite{Djordjevic:2011dd}, finite magnetic mass can be introduced
into Eq.~\eqref{col_rate_2}, reducing the collision rate to
\bea
\frac{\dif \Gamma_{e l}}{\dif^2 q}=\frac{C_A}{\pi} T \alpha(E T) \frac{\mu_E^2-\mu_M^2}{(q^2+\mu_E^2)(q^2+\mu_M^2)}, \label{col_rate_3}
\eea
where $\mu_M$ is the magnetic mass defined in the previous subsection, and
$C_A = 4/3$. This expression (i.e. Eq.~\eqref{col_rate_3}) can be further
reduced to
\bea
\frac{\dif \Gamma_{e l}}{\dif^2 q}=\frac{C_A}{\pi} T \alpha(E T)\left(\frac{1}{q^2+\mu_M^2}-\frac{1}{q^2+\mu_E^2}\right). \label{col_rate_4}
\eea

In the fluid rest frame, the transport coefficient $\hat{q}$ can then be computed as~\cite{JET:2013cls,Rapp:2018qla}
\bea
\hat{q} & =& \int_0^{\sqrt{6 E T}} \dif^2 q \, q^2 \cdot \frac{\dif \Gamma_{e l}}{\dif^2 q}\nonumber \\
& =&C_A T \alpha(E T) \int_0^{6 E T}  \dif q^2 \, q^2 \left(\frac{1}{q^2+\mu_M^2}-\frac{1}{q^2+\mu_E^2}\right)\nonumber\\
&=& C_A T \frac{4\pi}{(11-\frac{2}{3}n_f  )}
\frac{\left(\mu_E^2 \ln \left[\frac{6 E T+\mu_E^2}{\mu E^2}\right]-\mu_M^2 \ln\left[ \frac{6 E T+\mu_M^2}{\mu_M^2}\right]\right)}{\ln{(\frac{E T}{\Lambda^2})}} \label{q_hat}.
\eea
In the limit $ET \rightarrow \infty$, Eq.~(\ref{q_hat}) reduces to the expression independent of jet $E$:
\bea
\hat{q}&=&C_A T \frac{4 \pi}{11-\frac{2}{3} n_F}\left(\mu_E^2 \frac{\ln \frac{E T} { \mu_E^2 /6}}{\ln\frac{ E T}{  \Lambda^2}}-\mu_M^2 \frac{\ln \frac{E T} {\mu_M^2 / 6}}{\ln \frac{E T }{ \Lambda^2}}\right)  \nonumber\\
&\approx& C_A T \frac{4 \pi}{11-\frac{2}{3} n_F} \left(\mu_E^2-\mu_M^2\right) \nonumber\\
&=& C_A T \frac{4 \pi}{11-\frac{2}{3} n_F}  \frac{1+\frac{n_F}{6}}{11-\frac{2}{3} n_F} \frac{(4 \pi)^2 T^2}{W(\xi(T))}\left(1-x_{ME}^2\right) \nonumber\\
&=& C_A \left(\frac{4 \pi}{11-\frac{2}{3} n_F}\right)^2 \frac{4 \pi\left(1+\frac{n_F}{6}\right)}{W(\xi(T))} \left(1-x_{ME}^2\right) T^3, \label{q_hat_limit}
\eea
where $x_{ME}=\mu_M/\mu_E$ is the magnetic-to-electric mass ratio. It
is worth noticing that this is expected behavior: as a property of
the medium $\hat q$ should be independent (or weakly dependent) on jet
energy~\cite{JET:2013cls}. Nevertheless, many models/approaches fail to
describe this behavior~\cite{JET:2013cls}.

\section{Results}

\subsection{Constraining $\eta/s$ through high-$p_\perp$ data}

\begin{center}
\begin{figure}[tbh!]
 \begin{center}
 \includegraphics[scale=0.3]{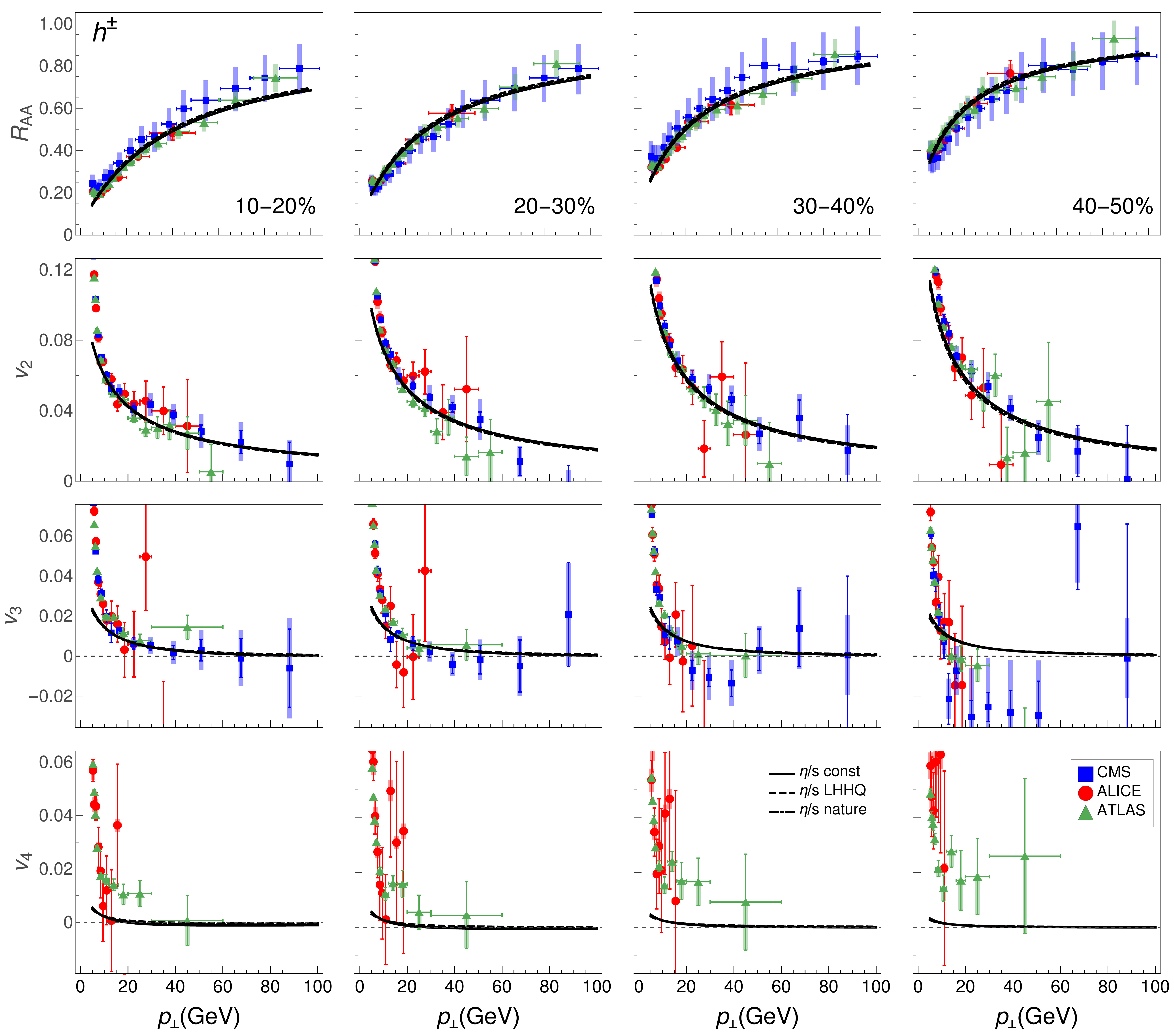}
 \caption{Charged hadron $R_{AA}$ (first row) and high-$p_\perp$ flow
   harmonics $v_2$ (second row), $v_3$ (third row) and $v_4$ (fourth
   row) as a function of transverse momentum in Pb+Pb collisions at
   $\sqrt{\sNN}=5.02$ TeV for different $(\eta/s)(T)$ parametrizations
   indicated in the legend.
   CMS (blue squares)~\cite{CMS:2017xgk,CMS:2016xef}, ALICE
   (red circles)~\cite{ALICE:2018vuu,ALICE:2018rtz}, and ATLAS (green
   triangles)~\cite{ATLAS:2018ezv,ATLAS:2017rmz} experimental data are
   also shown for comparison. Columns 1-4 represent the centrality
   classes 10-20\%, 20-30\%, 30-40\%, and 40-50\%, respectively.}
  \label{PbPb_ch_plot}
 \end{center}
\end{figure}
\end{center}

To examine the sensitivity of the high-$p_\perp$ observables on the
specific shear viscosity of the medium, we compare in
Fig.~\ref{PbPb_ch_plot} the experimental charged hadron $R_{AA}$  and
high-$p_\perp$ flow harmonics $v_2$, $v_3$ and $v_4$ in Pb+Pb
collisions at $\sqrt{\sNN}=5.02$ TeV to the theoretical predictions
calculated using three different $(\eta/s)(T)$ parametrizations (see
Sect.~\ref{hydro}). The high-$p_\perp$ flow harmonics are computed
using the scalar product method~\cite{Zigic:2022xks}. As seen in all
three cases, the calculated charged hadron $R_{AA}$ and flow anisotropies
are almost indistinguishable from each other.
Furthermore, the predicted charged hadron $v_4$ significantly
underestimates the experimental data even when the current large
experimental uncertainties are taken into account. We have previously
reported similar observation in Ref.~\cite{Zigic:2022xks} where
high-$p_\perp$ $v_4$ was calculated using several different
initializations of the fluid dynamical evolution.

\begin{center}
\begin{figure}[tbh!]
 \begin{center}
\includegraphics[scale=0.3]{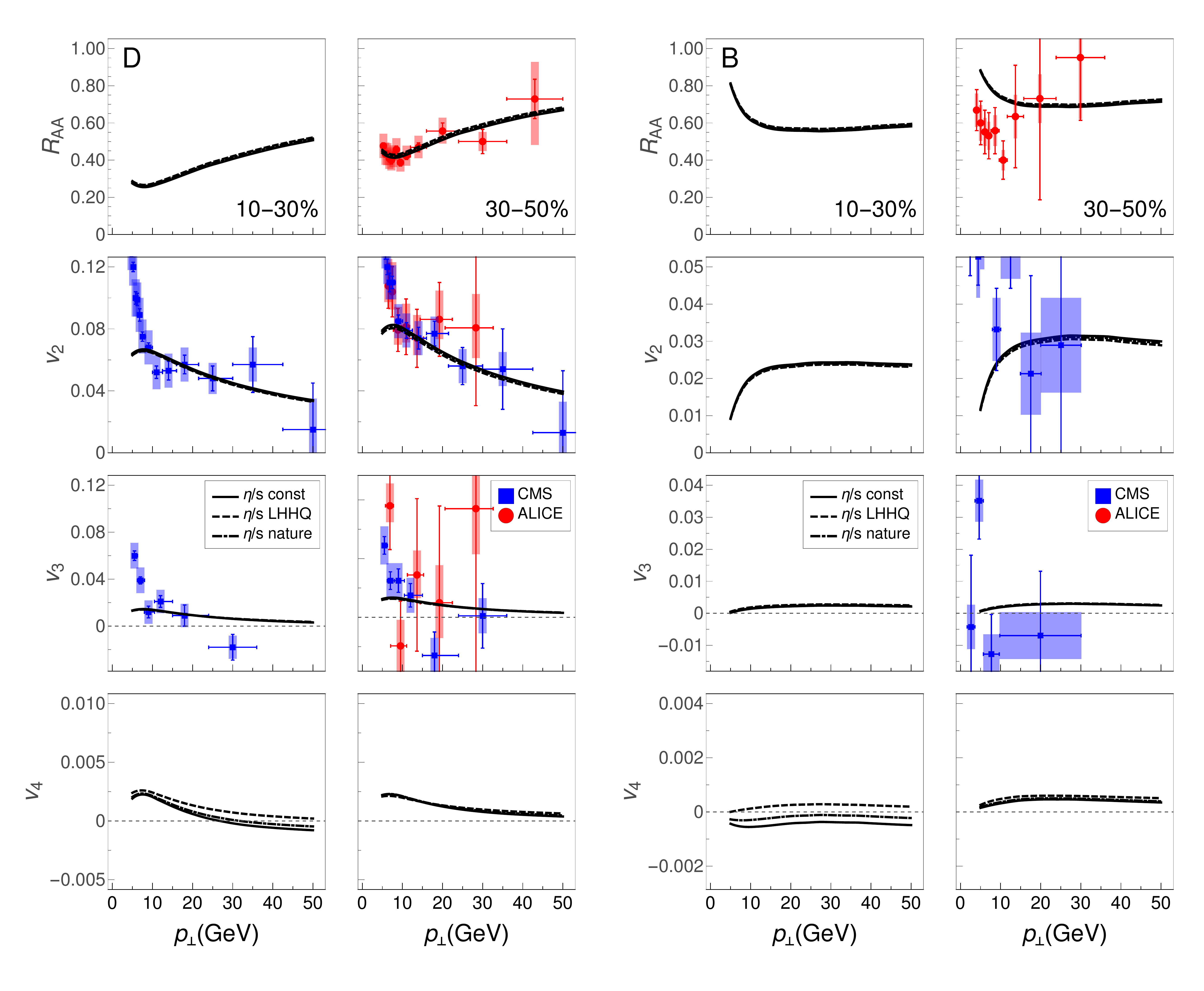}
 \caption{Predictions for D (left 4$\times$2 panel) and B meson (right
   4$\times$2 panel)$R_{AA}$ (first row) and high-$p_\perp$ flow
   harmonics $v_2$ (second row), $v_3$ (third row) and $v_4$ (fourth
   row) using three different $(\eta/s)(T)$ parametrizations at
   various centralities in Pb+Pb collisions at $\sqrt{\sNN}=5.02$ TeV.
   The theoretical predictions for D mesons are compared with the CMS
   (blue squares)~\cite{CMS:2020bnz} and ALICE (red
   circles)~\cite{ALICE:2021rxa, ALICE:2020iug} data, whereas B meson
   predictions are compared to preliminary CMS (blue
   squares)~\cite{CMS_B_meson_PbPb} and preliminary ALICE (red
   circles)~\cite{ALICE_B_meson_PbPb} data.}
  \label{PbPb_DB_plot}
 \end{center}
\end{figure}
\end{center}

Unfortunately, the heavy flavor high-$p_\perp$ observables shown in
Fig.~\ref{PbPb_DB_plot} are hardly more sensitive to the $(\eta/s)(T)$
parametrizations. The calculated D and B meson $R_{AA}$, $v_2$ and
$v_3$ in Pb+Pb collisions at $\sqrt{\sNN} = 5.02$ TeV do not depend on
our assumptions about $\eta/s$, whereas $v_4$ in the 10--30\%
centrality class shows some sensitivity. Nevertheless, given the large
experimental uncertainties of $v_2$ and $v_3$, it is doubtful whether
the small difference in $v_4$ is experimentally detectable, especially
when our $v_4$ predictions are very close to 0.

Since the collisions at LHC reach larger initial temperatures than
collisions at RHIC, we may expect them to be more sensitive to
$\eta/s$ at large temperatures, and thus to our $(\eta/s)(T)$
parametrizations. Nevertheless, for the sake of completeness and to
allow for surprises in the evolution, we checked whether the
high-$p_\perp$ observables measured in collisions at the full RHIC
energy ($\sqrt{\sNN} = 200$ GeV) allow us to distinguish between
different $(\eta/s)(T)$ parametrizations.

\begin{center}
\begin{figure}[tbh!]
 \begin{center}
 \includegraphics[scale=0.3]{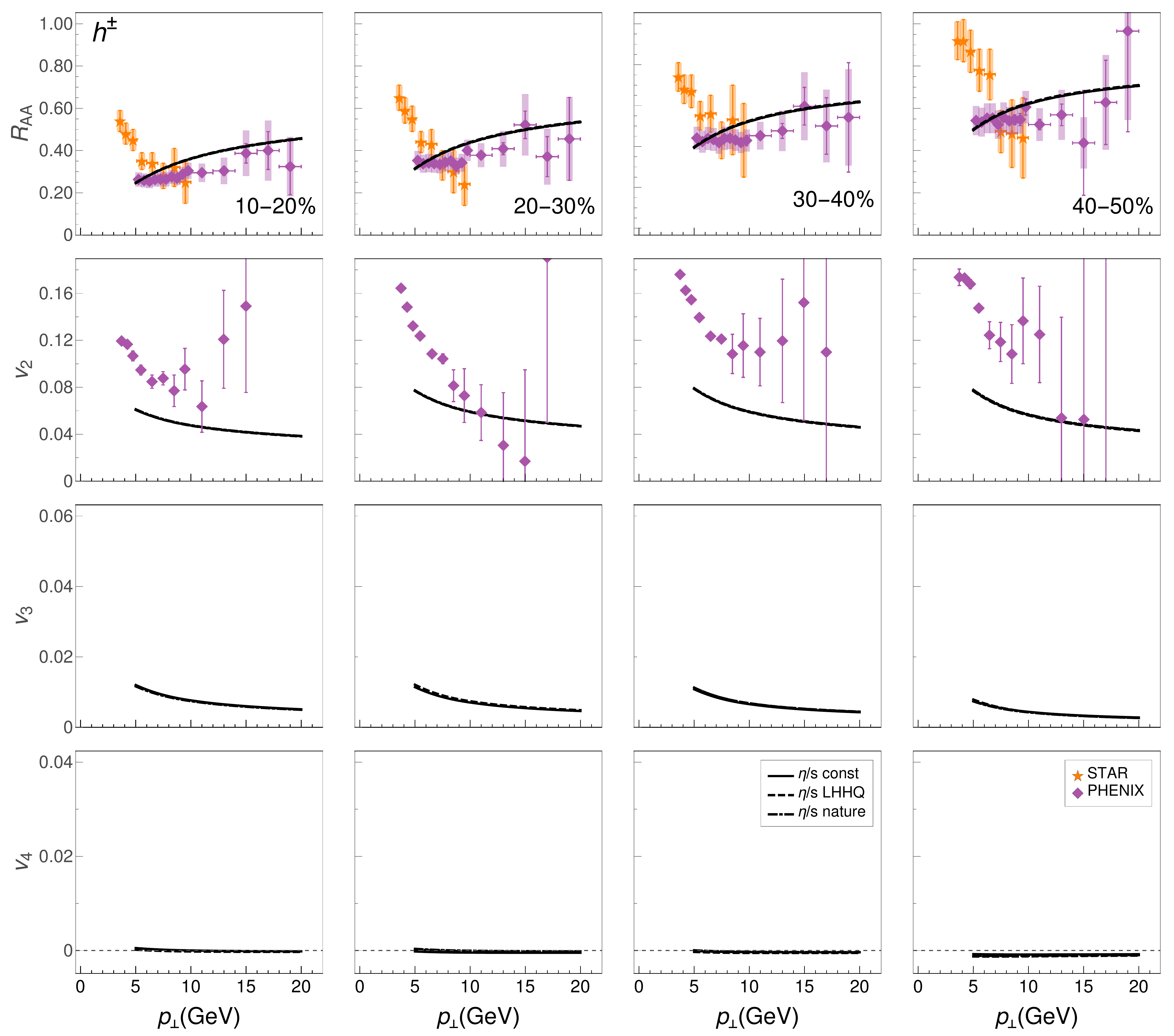}
 \caption{The calculated charged hadron $R_{AA}$ (first row),
   high-$p_\perp$ $v_2$ (second row), $v_3$ (third row) and $v_4$
   (fourth row) in $\sqrt{\sNN}=200$ GeV Au+Au collisions. The
   experimental data from the STAR (orange stars)~\cite{STAR:2003fka}
   and PHENIX (purple diamonds)~\cite{PHENIX:2012jha, PHENIX:2018wex}
   collaborations are also shown. Columns 1-4 represent the centrality
   classes 10-20\%, 20-30\%, 30-40\%, and 40-50\%, respectively.}
  \label{AuAu_ch_plot}
 \end{center}
\end{figure}
\end{center}

\begin{center}
\begin{figure}[tbh!]
 \begin{center}
 \includegraphics[scale=0.3]{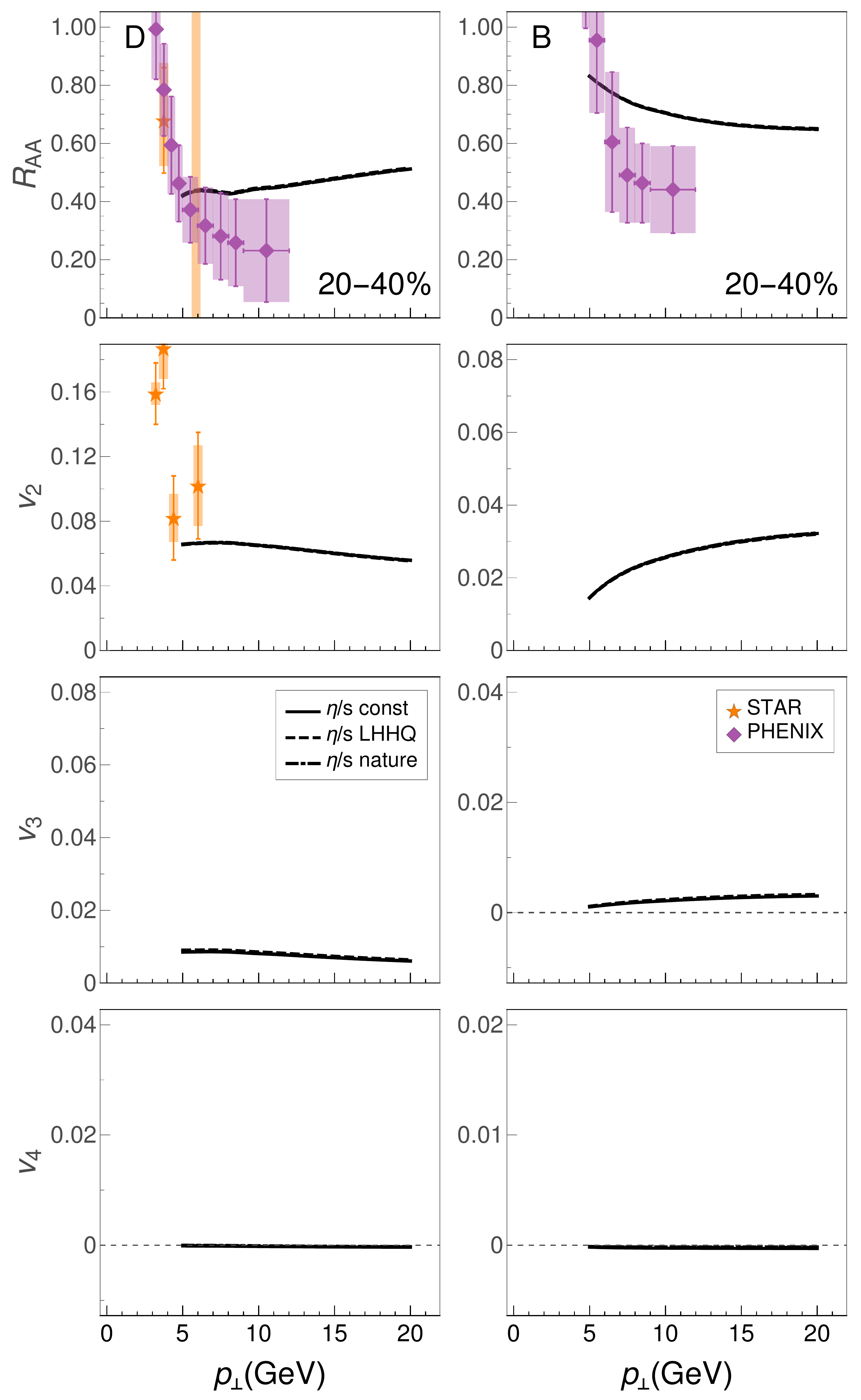}
 \caption{ D (left panel) and B meson (right panel) $R_{AA}$ (first
   row) and high-$p_\perp$ flow harmonics $v_2$ (second row), $v_3$
   (third row) and $v_4$ (fourth row) in $\sqrt{\sNN}=200$ GeV Au+Au
   collisions in 20-40\% centrality class. Theoretical predictions for
   D meson are compared with STAR (orange stars)~\cite{STAR:2014wif}
   and preliminary PHENIX (purple diamonds)~\cite{PHENIX_D_meson_AuAu}
   data, whereas B meson predictions are compared with the preliminary
   PHENIX (purple diamonds)~\cite{PHENIX_D_meson_AuAu} data.}
  \label{AuAu_DB_plot}
 \end{center}
\end{figure}
\end{center}

\begin{center}
\begin{figure}[tbh!]
 \begin{center}
 \includegraphics[scale=0.5]{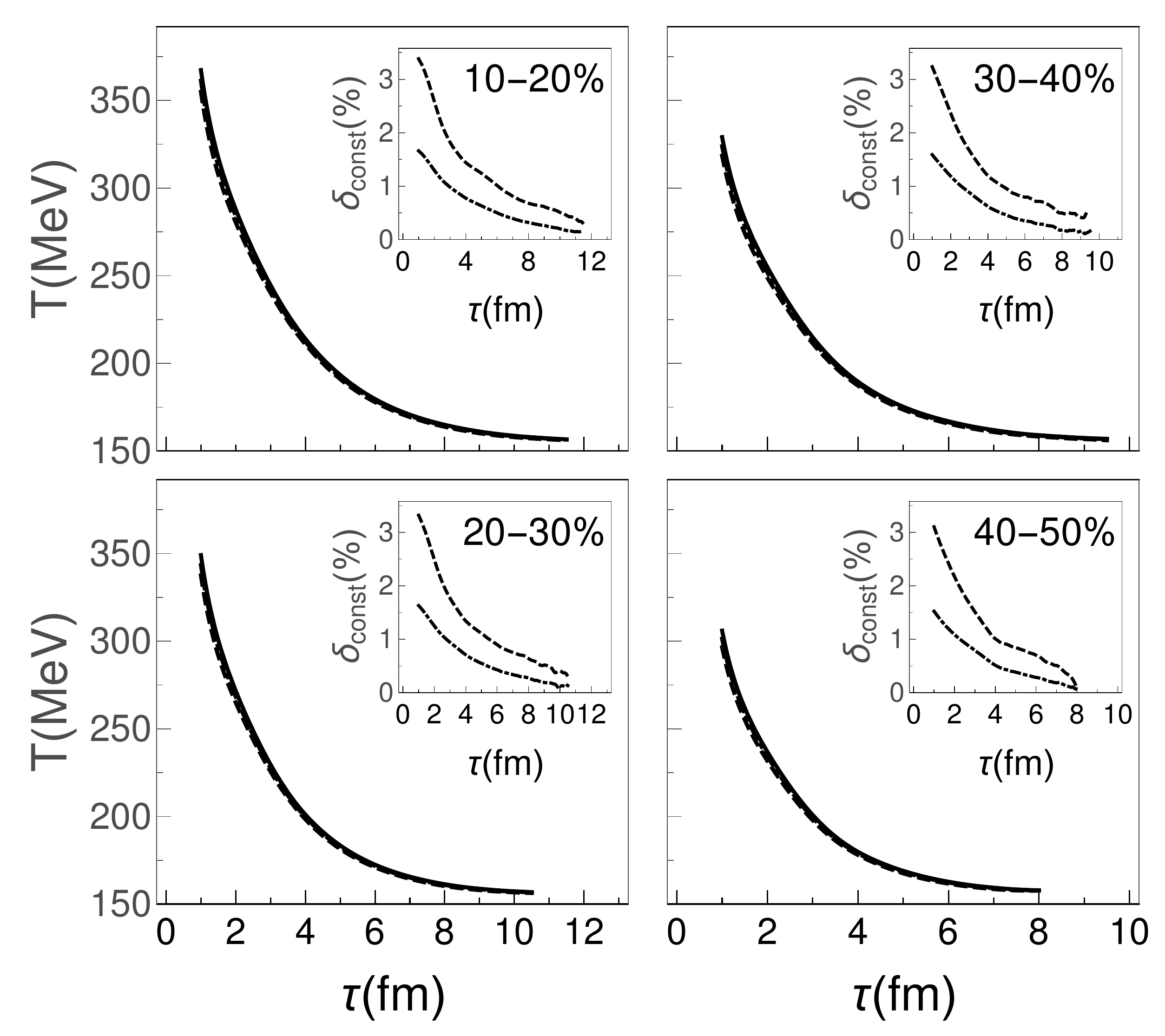}
 \caption{The average temperature experienced by the jets as a function
   of proper time for three different $(\eta/s)(T)$ parametrizations
   in $\sqrt{\sNN} = 5.02$ TeV Pb+Pb collisions for four different
   centrality regions (10-20\%, 20-30\%, 30-40\%, and 40-50\%, as
   indicated in each panel). The inset shows the relative difference
   in jet-perceived temperature in case of $`$Nature' (dot-dashed
   curve) and $`$LHHQ' (dashed curve) with
   respect to constant $\eta/s$ parametrizations. }
  \label{temp_tau_plot}
 \end{center}
\end{figure}
\end{center}

The theoretical predictions for charged hadron, and D and B meson
high-$\pT$ observables in Au+Au collisions at $\sqrt{\sNN} = 200$ GeV
collision energy are shown in Figs.~\ref{AuAu_ch_plot}
and~\ref{AuAu_DB_plot}, respectively. Again, we calculated our
predictions using the generalized DREENA-A framework with three
different $(\eta/s)(T)$ parametrizations. As can be seen, the
high-$p_\perp$ observables are not sensitive to the $\eta/s$ ratio at
high temperatures, and thus we cannot further constrain $(\eta/s)(T)$
using high-$p_\perp$ observables.

As was argued in the introduction, different $\eta/s$ require different initial temperatures. However, as shown in Fig.~\ref{temp_tau_plot}, temperature difference during the evolution is small and, as demonstrated above, insufficient to lead to observable differences in high-$p_\perp$ observables. In Fig.~\ref{temp_tau_plot}, we characterize the system temperature using the so-called average jet-perceived temperature: At each time $\tau$ we average the system temperature in
the transverse plane using the number of jets at each point as
weight - e.g., while the average initial temperature in (10-20)\% centrality class for Pb+Pb collisions at $\sqrt{s}=5.02\, TeV$ is 370 MeV, the maximal temperature experienced by the jet can reach up to 600 MeV. The jet-perceived temperatures, calculated using all three
$(\eta/s)(T)$ parametrizations are almost identical, differing less than 4\% at the early stages of the evolution and settling for less than 2\% for most of the evolution. The differences in the anisotropy of the jet-perceived temperature, $\langle jT_2\rangle$, introduced in Ref.~\cite{Stojku:2021yow}, are equally small (not shown). The investigated high-$p_\perp$ observables turned out to be insensitive
to such small differences in temperature.

Even if our calculated high-$\pT$ $R_{AA}$ and $v_2$ agree with the
data (see Fig.~\ref{PbPb_ch_plot}), the calculated $\langle jT_2\rangle$
are slightly below the experimentally favored values. This deviation
is possible since our results for both $R_{AA}$ and $v_2$ are at the
lower end of experimental uncertainty. When taking the ratio
$v_2/(1-R_{AA})$, which constrains $\langle jT_2\rangle$, this deviation
from the data is magnified, and the calculated $\langle jT_2\rangle$
is below the experimental constraint. Nevertheless the values of
$\langle jT_2\rangle$ obtained in these calculations are close to the
largest values obtained in Ref.~\cite{Stojku:2021yow} for various
initialization models.

\subsection{Calculating $\eta/s$ from the dynamical energy loss $\hat{q}$}

In our previous publications, we have seen that the DREENA framework is
capable of reproducing the observed $R_{AA}$ without fitting
parameters~\cite{Djordjevic:2013xoa,DDB_PLB,DREENA_C} (see also comparison to $R_{AA}$
in the previous subsection). This agreement suggests that the
dynamical energy loss formalism can adequately describe interactions
between high-$p_\perp$ particles and the QCD medium. Thus, it seems
reasonable to estimate $(\eta/s)(T)$ theoretically using the
dynamical energy loss model.

For this purpose, we need to estimate the jet quenching parameter $\hat{q}$, quantifying the transverse momentum broadening of fast parton due to its elastic scatterings with the medium~\cite{BDMPS}. This parameter is a key quantity in estimating the interaction strength between jet partons and nuclear matter~\cite{Guo:2000nz,Majumder:2009ge,JET:2013cls,Barata:2020rdn,Sirimanna:2021sqx,Grishmanovskii:2022tpb}. It has been proposed to be a valuable tool for various purposes, including gaining insights into the jet quenching phenomenon, estimating the bulk medium property $(\eta/s)(T)$~\cite{MBW,Muller}, and more recently, exploring the QCD phase diagram~\cite{Wu:2022vbu}.

We presented the derivation of the transport coefficient
$\hat q$ from our dynamical energy loss model in
section~\ref{derive-qhat}.
We note that $\hat{q}$ is weakly dependent on $E$ due to $\ln(ET)$
appearing both in the numerator and the denominator of
Eq.~\eqref{q_hat}, as desired for a medium property such as transport
coefficient.

\begin{center}
\begin{figure}[tbh!]
 \includegraphics[scale=0.55]{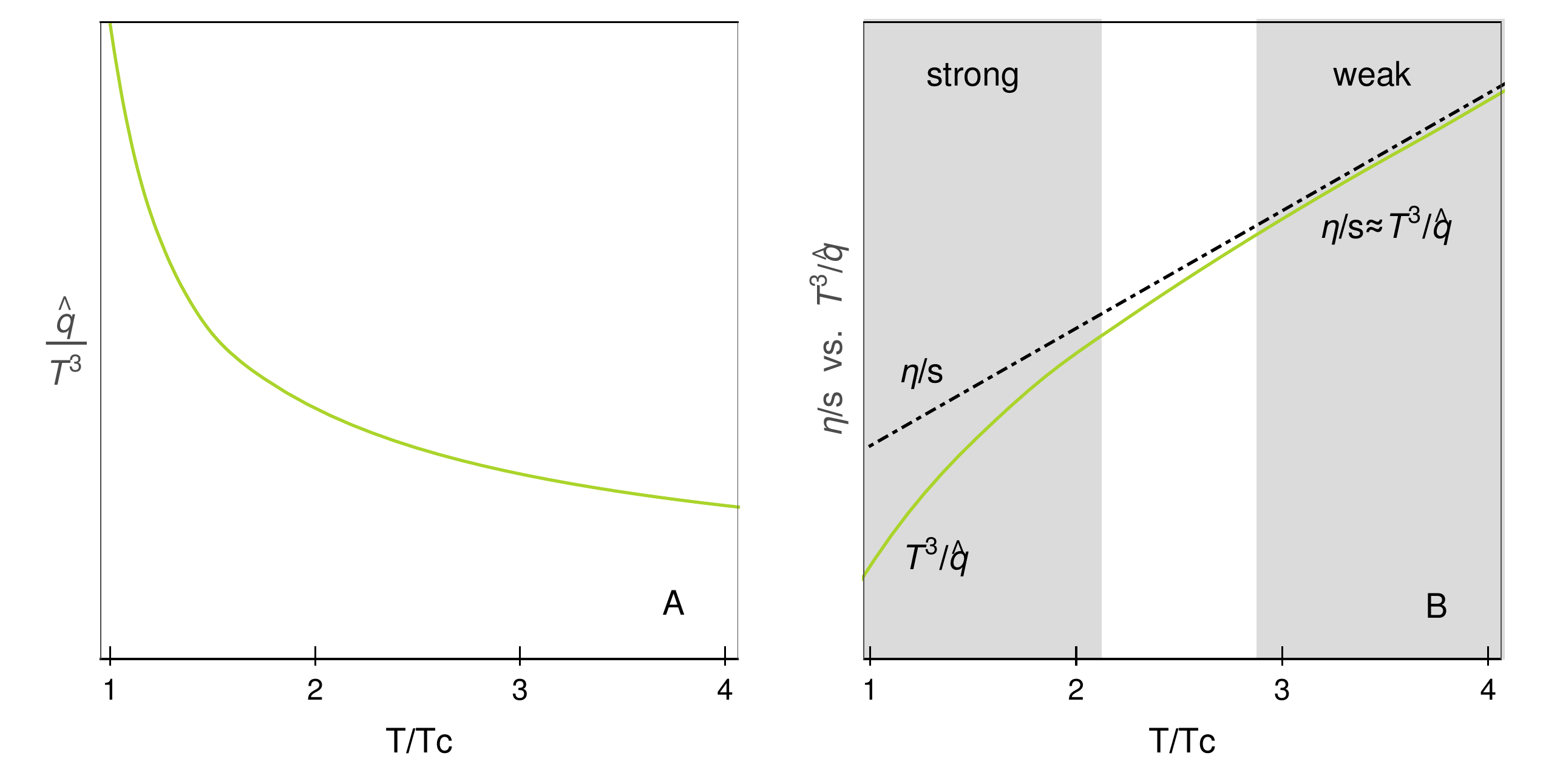}
 \caption{({\bf A}) A schematic $T$ dependence of quenching strength
   $\hat{ q} / T^3$ proposed in~\cite{Liao}. ({\bf B}) A scheme for
   mapping soft-to-hard boundary based on Ref.~\cite{MBW}.}
   \label{schematic}
\end{figure}
\end{center}

Before discussing our results further, we outline the theoretical
expectations for $\hat q$ and its relationship to $\eta/s$.
%As is well known, $\eta$ and $\hat{q}$ are two important QGP transport
%coefficients.
To account for the temperature dependence of the coefficients $\eta$
and $\hat q$, it is common practice to examine their dimensionless
counterparts: $\eta/s$, and $\hat q/T^3$~\cite{Muller}. Both
quantities are sensitive to the effective coupling strength in QGP. If
the coupling is weak, $\eta/s$ is large, while $\hat{q}/T^3$ is
small. Conversely, when the coupling is strong, $\eta/s$ becomes
small, while $\hat{q}/T^3$ is large. In the case of weak coupling, it
has been argued that these two quantities are related by $
\frac{\eta}{s} \cdot \frac{\hat{q}}{T^3} \approx \mathrm{const}$,
i.e., more specifically~\cite{MBW,Muller}:
\bea
\frac{\eta}{s} \approx 1.25 \,\frac{T^3}{\hat{q}}, \label{etas_qhat}
\eea

Furthermore, to explain the large observed high-$p_\perp$ $v_2$, it
was proposed in Ref.~\cite{Liao}, that the jet-quenching factor
$\hat{q}/T^3$ must rise rapidly when approaching $T_c$ from
above. We {\it schematically} depicted such behavior in
Fig.~\ref{schematic}A. A behavior that is not straightforward nor
trivial to obtain from a model calculation.

The expected (qualitative) relation of the $T^3/\hat{q}$ and $(\eta/s)(T)$, -- based on the existing knowledge from previous studies -- is {\it schematically} depicted in Fig.~\ref{schematic}B~\cite{MBW}. At large temperatures, we expect the system to be weakly coupled. At that limit, our dynamical energy loss
model should be applicable, and Eq.~(\ref{etas_qhat}) should be a good
approximation. Thus, we expect the calculated $T^3/\hat{q}$ to agree
well with the inferred $\eta/s$ as shown in the grey area at the right
part of Fig.~\ref{schematic}B. On the other hand, in the strongly
coupled limit (the left gray area in Fig.~\ref{schematic}B) close to
$T_c$, the calculated $T^3/\hat{q}$ is expected to significantly
deviate from the inferred $\eta/s$. Interestingly, the $T^3/\hat{q}$
calculated using weak coupling methods is expected to drop below the
inferred $\eta/s$~\cite{MBW}, which, as known, is very close to the
AdS/CFT lower limit of $1/(4\pi)$ in the vicinity of $T_c$.

The region between strongly and weakly coupled limits is the so-called
$``$soft-to-hard" boundary~\cite{SHBoundary}, i.e., the region where
the transition from a strongly to a weakly coupled regime could take
place. Therefore, plotting together $\eta/s$ and $T^3/ \hat{ q}$ as a
function of $T$ might allow estimating the $``$soft-to-hard" boundary,
as the region where these two curves start to deviate, as
schematically shown in Fig.~\ref{schematic}B.

\begin{center}
\begin{figure}[tbh!]
 \includegraphics[scale=0.55]{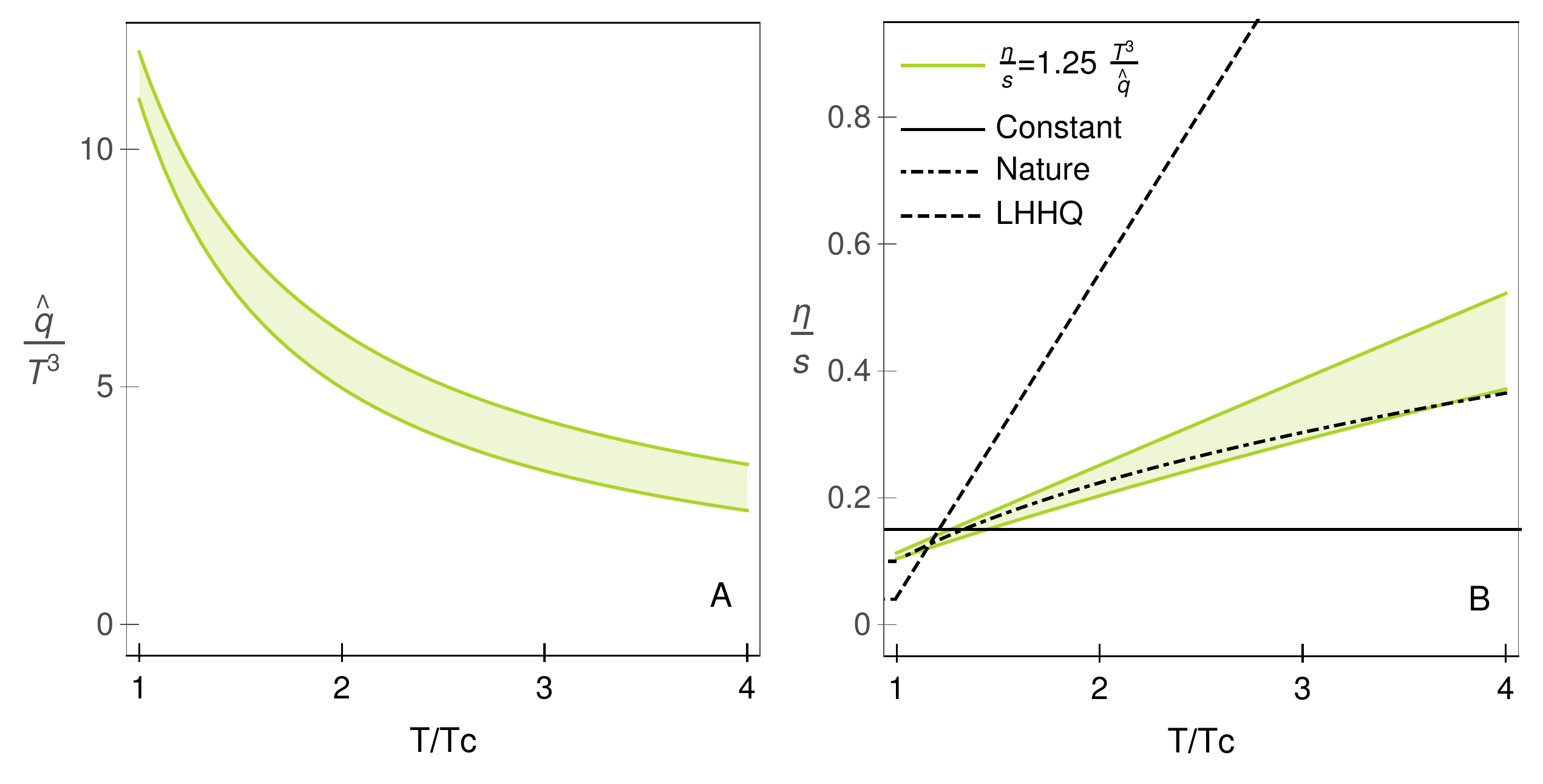}
 \caption{({\bf A}) $T$ dependence of $\hat{ q} / T^3$ extracted from
   dynamical energy loss for initial jet energy in the range $E=3~GeV$
   (lower boundary) to $E=10~GeV$ (upper boundary)~\cite{JET:2013cls}.
   ({\bf B}) Comparison of $\eta/s$ extracted from $\hat{ q} / T^3$
   (shown in A), with three different choices of the specific shear
   viscosity considered in this study and indicated in the legend.}
  \label{Panel}
\end{figure}
\end{center}

We calculate $\hat{q}/T^3$ from our dynamical energy loss using
Eq.~\eqref{q_hat} in the initial jet energy range
$3~\mathrm{GeV} < E < 10~\mathrm{GeV} $, as $p_\perp$ has to be low enough to mimic
interactions of partons within the medium. The obtained result is shown in
Fig.~\ref{Panel}A, and qualitatively similar to the expectation shown
in Fig.~\ref{schematic}A.  In particular, near $T_c$, we obtain an enhanced quenching, which is considerably larger than quenching in other energy loss models~\cite{JET:2013cls} (with the exception of~\cite{Grishmanovskii:2022tpb}, which got a substantial increase in $\hat{q}/T^3$ near $T_c$, due to a very large coupling in their model). Some models
even predicted a decrease of $\hat{q}/T^3$ (or increase in $\eta/s$)
when the temperature is approaching $T_c$ from
above~\cite{MWL,MBCAB,JET:2013cls}. The enhancement near $T_c$, obtained in Fig.~\ref{Panel}A, is due
to an interplay between chromo-electric and chromo-magnetic
screenings~\cite{Djordjevic:2011dd}. As the magnetic component is
inherently related to the dynamical nature of the medium constituents,
it cannot exist in widely used static models, making the evolving
medium an important feature of the dynamical energy loss model.

We convert our calculated $\hat{q}/T^3$ to $(\eta/s)(T)$ using
Eq.~\eqref{etas_qhat}, and compare it to the parametrizations used in
this study in Fig.~\ref{Panel}B. First, the uncertainty due to the
relevant initial jet energy is way smaller than the range of our
$(\eta/s)(T)$ parametrizations. Second, our result is surprisingly
close to the parametrization inspired by the Bayesian analysis of
Ref.~\cite{Bernhard:2019bmu}, ``Nature'', and, third, unlike expected,
our result obtained using weak coupling approximation does not drop
significantly below the inferred $\eta/s$ values in the vicinity of
$T_c$.

\begin{center}
  \begin{figure}[tbh!]
 \includegraphics[scale=0.8]{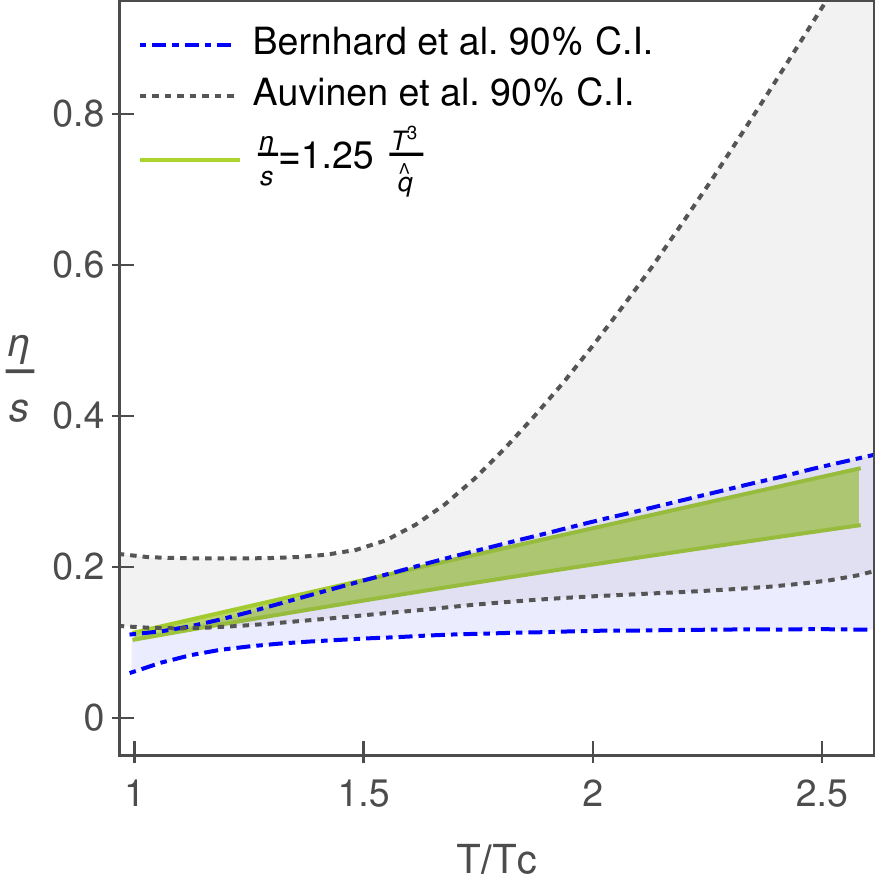}
 \caption{ Comparison of $\eta/s$ extracted from
   $\hat{q}/T^3$ to the 90\% credible intervals of the Bayesian
   analyses of Refs.~\cite{Bernhard:2019bmu,Auvinen:2020mpc} (Bernhard
   \textit{et al.} and Auvinen \textit{et al.}, respectively).}
  \label{Bayes-comparison}
\end{figure}
\end{center}

To further gauge the significance of our result, we compare it to the
90\% credible intervals for $(\eta/s)(T)$ obtained in two
state-of-the-art Bayesian analyses~\cite{Bernhard:2019bmu,
  Auvinen:2020mpc} in Fig.~\ref{Bayes-comparison}. Interestingly, the
$\eta/s$ dependence extracted from our $\hat{q}/ T^3$ shows an
excellent agreement with both of these analyses in the entire $T$
range, i.e., it falls precisely in the overlap of the two intervals.
Not only does our result agree at large temperatures, where the
Bayesian constraints are weakest, but even in the vicinity of $T_c$,
where we expected our result to drop below the inferred values of
$\eta/s$ (as depicted in Fig.~\ref{schematic}B).  This is a surprising
result, as one might expect that our calculation of $\hat{q}$ from the
dynamical energy loss model and Eq.~\eqref{etas_qhat} are reliable
only in the weakly coupled regime. However, the agreement extends to
$T_c$, i.e., to the regime corresponding to strong coupling.

While the extended agreement observed in Fig.~\ref{Bayes-comparison}
is encouraging in terms of the prediction ability of the dynamical
energy loss formalism, it leads to the question of why the expected
behavior (shown schematically in Fig.~\ref{schematic}B) is not
observed in Fig.~\ref{Bayes-comparison}. It is unlikely that the weak
coupling regime would extend down to $T_c$. Instead, it was
suggested~\cite{MBW} that Eq.~\eqref{etas_qhat} is valid as long as
the quasiparticle picture of QGP is applicable. The same is required
for the validity of energy loss calculations, including our dynamical
energy loss model. Therefore, it is an intriguing (and potentially
significant) hypothesis that the quasiparticle picture used for
describing interactions between jet and QGP is consistent with the QCD
medium created at RHIC and LHC at the entire temperature range. In
practical terms, this hypothesis is consistent with the dynamical
energy loss model's ability to explain a wide range of experimental
data.

Lastly, one of the open issues of QGP physics is mapping the
$``$soft-to-hard" boundary. As discussed, a possible approach for
estimating the boundary is to compare estimates of the same quantity
(like $\eta/s$) from the high-$p_\perp$ and low-$p_\perp$ sector as
schematically presented in Fig.~\ref{schematic}B. However, as shown
here, the $\eta/s$ obtained from high-$p_\perp$ theory and inferred
from the low-$p_\perp$ data agree in the entire $T$ range providing no
guidance on locating the boundary.

\section{Summary}

Our previous studies showed that combining high-$p_\perp$
predictions/data with temperature profiles from bulk medium
simulations can constrain QGP medium properties, such as its early
evolution and medium averaged anisotropy. Here we used an equivalent
approach, where temperature profiles corresponding to different
$(\eta/s)(T)$ parametrizations were generated and subsequently used by
our generalized DREENA-A framework to generate predictions for $R_{AA}$ and
high-$p_\perp$ $v_2$, $v_3$ and $v_4$. However, we found that this
approach cannot differentiate between temperature profiles generated
using different $(\eta/s)(T)$ parametrizations, since the differences
in $T$ profiles and jet-perceived anisotropies~\cite{Stojku:2021yow}
turned out to be small, and consequently the differences in
high-$p_\perp$ predictions were also small. It is unrealistic to
expect that the experiments at RHIC and LHC will (in a reasonable time
frame) achieve the precision needed to distinguish between these
predictions.

On the other hand, our second approach, based on calculating the
quenching strength $\hat{q}/T^3$ from our dynamical energy loss model,
showed a surprisingly good agreement with the constraints to
$(\eta/s)(T)$ extracted from low-$p_\perp$ data by state-of-the-art
Bayesian analysis. Such agreement is highly nontrivial as it
originates from two entirely different approaches: a theoretical
calculation based on finite temperature field theory through
generalized HTL approach (dynamical energy loss) and inferring
$(\eta/s)(T)$ from experimental data using fluid-dynamical modeling
and advanced statistical (Bayesian) methods. The agreement is also
surprising, as it extends all the way to $T_c$, where a strongly
coupled regime should apply, and where a disagreement between energy
loss calculation based on weak coupling approximation and inferred
value of $\eta/s$ is expected. We interpret the absence of such a
disagreement in terms of the quasiparticle picture being valid even
close to $T_c$. However, this obscures estimating soft-to-hard
boundary, whose inference remains one of the field's major (to our
knowledge, unresolved) problems. Overall, this work further emphasizes
the utility of jet tomography, where low- and high-$p_\perp$ theory
and data are jointly used to constrain the QGP properties.

{\em Acknowledgments:}
This work is supported by the European Research Council, grant
ERC-2016-COG: 725741, and by the Ministry of Science and Technological
Development of the Republic of Serbia. PH was also supported by the
program Excellence Initiative--Research University of the University of
Wroc\l{}aw of the Ministry of Education and Science.
JA acknowledges the financial support from the Academy of Finland Project
No. 330448. JA's research was also funded as a part of the Center of
Excellence in Quark Matter of the Academy of Finland (Project No.~346325).
This research is part of the European Research Council Project No.\
ERC-2018-ADG-835105 YoctoLHC.

\end{document}